# Theoretical Insights into the Topology of Molecular Excitons from Single-Reference Excited States Calculation Methods

Thibaud Etienne

Additional information is available at the end of the chapter

**Abstract**

This chapter gives an introduction to qualitative and quantitative topological analyses of molecular electronic transitions. Among the possibilities for qualitatively describing how the electronic structure of a molecule is reorganized upon light absorption, we chose to detail two of them, namely, the detachment/attachment density matrix analysis and the natural transition orbitals strategy. While these tools are often introduced separately, we decided to formally detail the connection existing between the two paradigms in the case of excited states calculation methods expressing any excited state as a linear combination of singly excited Slater determinants, written based on a single-reference ground state wave function. In this context, we show how the molecular exciton wave function plays a central role in the topological analysis of the electronic transition process.

**Keywords:** excited states, excitons, detachment/attachment, transition matrix and orbitals, charge transfer

## 1. Introduction

Providing a quantitative insight into light-induced electronic structure reorganization of complex chromophores remains a challenging task that has attracted a substantial attention from theoretical communities in the past few years [1–15]. Indeed, a potential knowledge related to the ability of a chromophore to undergo a charge transfer caused by photon absorption or emission [16, 17] is of seminal importance for designing novel dyes with highly competitive optoelectronic properties [18–21]. Most often, such quantitative probing of the charge transfer locality is accompanied by a qualitative study of the rearrangement of the electronic distribution in the molecule, and the aim of this contribution is to demonstrate how in certain cases different topological paradigms are formally connected, with the junction point being the definition of the molecular exciton wave function.

The outcome of the computation of the molecular electronic excited states using a quantum calculation method is, in addition to the transition energy, a series of mathematical objects allowing one to analyze the transition topology. If the reference ground state wave function is, in a given basis (called the canonical basis), written as a Slater determinant, any excited state written based on this ground state wave function is called a single-reference excited state. From this single reference and in a given canonical basis, some methods express excited states as a linear combination of singly excited Slater determinants, which means that the excited state wave function is written as a pondered sum of Slater determinants constructed from the ground state reference, in which one occupied spinorbital (*vide infra*) is replaced by a virtual one. This type of excited state construction is often referred to as a configuration interaction (CI) solely involving singly excited Slater determinants. In our case, the reference ground state wave function can be a Hartree-Fock or a Kohn-Sham Slater determinant, and the excited states calculation methods we deal with in this paper are called configuration interaction singles (CIS), time-dependent Hartree-Fock (TDHF), random-phase approximation (RPA), Tamm-Dancoff approximation (TDA), or time-dependent density functional theory (TDDFT). For more details about the machinery of these methods, see Refs. [22–25]. While in the case of CIS and TDA, the determination of the exciton wave function is very straightforward, for the other methods, it has been subject to the so-called assignment problem which consisted in providing a CI structure to the TDDFT excited state (since the central RPA/TDHF and TDDFT equations have the same structure, the assignment problem is transferable to these methods also) [26, 27].

Based on the outcome of the excited states calculation, one can select an electronic transition of interest and inspect the different hole/particle contributions from the occupied/virtual canonical subspaces for having an insight into the light-induced charge displacement topology. However, in some occurrences, such analyses are quite cumbersome because many of these contributions can be significant while bearing a divergent physical meaning. For the purpose of providing a straightforward picture of the electronic transition topology, multiple tools were developed. Among them, one can cite the detachment/attachment strategy [3, 4, 25, 28–31], which delivers a one-electron charge density function for the hole and for the particle that are generated by photon absorption. This strategy is based on the diagonalization of the so-called difference density matrix (the difference between the excited and ground state density matrices) and a sorting of the resulting "transition occupation numbers" based on their sign. The result of this analysis is a simple identification of the photogenerated depletion and increment zones of charge density. Quantitative insights are then reachable through the manipulation of the detachment/attachment density functions and the definition of quantum metrics [3–5]. On the other hand, one can consider the projection of the exciton wave function in the canonical basis through the so-called transition density matrix [13, 25, 29, 30, 32–43], which singular value decomposition [44] provides the most compact spinorbital representation of the electronic transition. The great advantage of this method is that in most of the cases it condensates the physics of an electronic transition into one couple of hole/particle wave functions.

This chapter first recalls some useful concepts related to the reduced density matrix formalism and its relation to the notion of electron density and density matrix in a canonical space. The detachment/attachment density matrix construction is then exposed in details and is used for quantifying the charge transfer locality through several quantum descriptors. Afterward, the notion of density matrix is extended to electronic transitions through the concept of transition density matrix. The information contained in this particular matrix is shown to be extractable and is discussed in details by introducing the so-called natural transition orbitals. The detachment/attachment and natural transition orbitals formalisms are then compared, and we demonstrate that the difference density matrix is constructed from the direct sum of two matrix products involving only the transition density matrix, that is, the molecular exciton wave function projected into the canonical space (Lemma III.1). It follows that the natural transition orbitals are nothing but the eigenvectors of the detachment/attachment density matrices (Theorem III.1), which is a major conclusion in this contribution since the two formalisms are often introduced as being distinct and belonging to two separate paradigms. This conclusion is finally used for showing that the quantum indices designed for quantifying the charge transfer range and magnitude can be equivalently derived from the detachment/attachment and natural transition orbitals paradigms (Corollary III.1).

All the derivations are performed in the canonical space in the main text, but the important concepts and conclusions are also written in the basis of atomic functions in Appendix B. The calculations performed for this book chapter were done using the G09 software suite [45].

## 2. Theoretical background

Since this chapter will be mostly dealing with quantum state density matrices, the first paragraph of this section consists in a short reminder about the one-particle reduced density matrices corresponding to single-determinant wave functions.

### 2.1. One-particle reduced density matrix

We consider an $N$–electron system, with the $N$ electrons being distributed in $L$ spinorbitals ($N$ occupied, $L-N$ virtual). In this contribution we will write any ground state wave function $\psi_0$ as an arrangement of the occupied spinorbitals into a single Slater determinant. The density matrix kernel representing the corresponding ground electronic state writes

$$\tilde{\gamma}^0(\mathbf{r}_1, \mathbf{r}_1') = N \sum_{\sigma_1 = \alpha, \beta} \int d\mathbf{x}_2 \ldots \int d\mathbf{x}_N \, \psi_0(\mathbf{r}_1, \sigma_1, \ldots, \mathbf{x}_N) \psi_0^*(\mathbf{r}_1', \sigma_1, \ldots, \mathbf{x}_N) = \sum_{r=1}^{L} \sum_{s=1}^{L} \varphi_r(\mathbf{r}_1) (\boldsymbol{\gamma}^0)_{rs} \varphi_s^*(\mathbf{r}_1'), \quad (1)$$

where $\mathbf{x}$ is a four-dimensional variable containing the spatial ($\mathbf{r}$) and the spin-projection ($\sigma$) coordinates. The density matrix kernel reduces to the electron density function when $\mathbf{r}_1 = \mathbf{r}_1'$, and its integral over the whole space returns the number of electrons:

$$\tilde{\gamma}^0(\mathbf{r}_1, \mathbf{r}_1) \equiv n_0(\mathbf{r}_1) = \sum_{r=1}^{L} \sum_{s=1}^{L} \varphi_r(\mathbf{r}_1) (\boldsymbol{\gamma}^0)_{rs} \varphi_s^*(\mathbf{r}_1) \Rightarrow \int_{\mathbb{R}^3} d\mathbf{r}_1 \, \tilde{\gamma}^0(\mathbf{r}_1, \mathbf{r}_1) = \int_{\mathbb{R}^3} d\mathbf{r}_1 \, n_0(\mathbf{r}_1) = N. \quad (2)$$

The $(\gamma^0)_{rs}$ terms appearing in Eq. (1) are the elements of the one-particle reduced density matrix expressed in the canonical space of spinorbitals $\{\varphi\}$ and can be isolated by integrating the product of $\tilde{\gamma}^0$ with the corresponding spinorbitals

$$\left(\gamma^0\right)_{rs} = \int_{\mathbb{R}^3} d\mathbf{r}_1 \int_{\mathbb{R}^3} d\mathbf{r}_1' \; \varphi_r^*(\mathbf{r}_1)\, \tilde{\gamma}^0\left(\mathbf{r}_1, \mathbf{r}_1'\right) \varphi_s\left(\mathbf{r}_1'\right). \tag{3}$$

Note that generally speaking the $r \times s$ density matrix element in a given spinorbitals space $\{\varphi\}$ for a given quantum state $|\psi\rangle$ writes

$$\left(\gamma\right)_{rs} = \left\langle \psi | \hat{r}^\dagger \hat{s} | \psi \right\rangle; \; \gamma \in \mathbb{R}^{L \times L} \tag{4}$$

where conventionally $r$ and $s$ indices range from 1 to $L$. In Eq. (4) we introduced the annihilation and creation operators from the second quantization.

## 2.2. Detachment and attachment density matrices

One known strategy for formally assigning the depletion and increment zones of charge density appearing upon light absorption is the so-called detachment/attachment formalism. This approach consists in separating the contributions related to light-induced charge removal and accumulation by diagonalizing the one-particle difference density matrix $\gamma^\Delta \in \mathbb{R}^{L \times L}$. Such matrix is obtained by taking the difference between the target excited state $|\psi_x\rangle$ and the ground state $|\psi_0\rangle$ density matrices:

$$\gamma^\Delta = \gamma^x - \gamma^0. \tag{5}$$

This density matrix can be projected into the Euclidean space in order to directly visualize the negative and positive contributions to the light-induced charge displacement:

$$n_\Delta(\mathbf{r}_1) = \sum_{r=1}^{L} \sum_{s=1}^{L} \varphi_r(\mathbf{r}_1) \left(\gamma^\Delta\right)_{rs} \varphi_s^*(\mathbf{r}_1) = n_x(\mathbf{r}_1) - n_0(\mathbf{r}_1). \tag{6}$$

Note that since no fraction of charge has been gained or lost during the electronic transition, the integral of this difference density over all the space is equal to zero:

$$\int_{\mathbb{R}^3} d\mathbf{r}_1 \; n_\Delta(\mathbf{r}_1) = \underbrace{\int_{\mathbb{R}^3} d\mathbf{r}_1 \; n_x(\mathbf{r}_1)}_{N} - \overbrace{\int_{\mathbb{R}^3} d\mathbf{r}_1 \; n_0(\mathbf{r}_1)}^{N} = 0. \tag{7}$$

However, visualizing this difference density does not provide a straightforward picture of the transition. The interpretation of the transition in terms of charge density depletion and increment can be made more compact by diagonalizing the difference density matrix:

$$\exists \mathbf{M} \; | \; \mathbf{M}^\dagger \gamma^\Delta \mathbf{M} = \mathbf{m} \tag{8}$$

where $\mathbf{m}$ is a diagonal matrix and $\mathbf{M}$ is unitary. Similar to the eigenvalues of a quantum state density matrix, the eigenvalues of $\gamma^\Delta$, contained in $\mathbf{m}$, can be regarded as the occupation numbers of the transition in the canonical space. Those can be negative or positive, corresponding, respectively, to charge removal or accumulation. These eigenvalues can therefore be sorted with respect to their sign:

$$\mathbf{k}_{\pm} = \frac{1}{2}\left(\sqrt{\mathbf{m}^2} \pm \mathbf{m}\right) \tag{9}$$

where $\mathbf{k}_+$ (respectively, $\mathbf{k}_-$) is a diagonal matrix storing the positive (absolute value of negative) eigenvalues of the difference density matrix. These two diagonal matrices can be separately backtransformed to provide the so-called detachment ($d$) and attachment ($a$) density matrices and the corresponding charge densities:

$$\mathbf{M}\mathbf{k}_-\mathbf{M}^\dagger = \gamma^d \xrightarrow{\mathbb{R}^3} n_d(\mathbf{r}) = \sum_{r=1}^{L}\sum_{s=1}^{L}\left(\gamma^d\right)_{rs}\ \varphi_r(\mathbf{r})\varphi_s^*(\mathbf{r}); \ \mathbf{M}\mathbf{k}_+\mathbf{M}^\dagger = \gamma^a \xrightarrow{\mathbb{R}^3} n_a(\mathbf{r}) = \sum_{r=1}^{L}\sum_{s=1}^{L}\left(\gamma^a\right)_{rs}\ \varphi_r(\mathbf{r})\varphi_s^*(\mathbf{r}). \tag{10}$$

These detachment/attachment densities ($n_d(\mathbf{r})$ and $n_a(\mathbf{r})$) are then nothing but the hole and particle densities we were seeking. These densities are reproduced in **Figure 1** for two paradigmatic cases of electronic transitions: one local transition and one long-range charge transfer. In the next paragraph, we will see how the locality of a charge transfer can be quantified using the detachment/attachment charge densities.

### 2.3. Quantifying the charge transfer locality

One possible strategy for evaluating the magnitude of the electronic structure reorganization is to compute the spatial overlap between the hole and the particle. This is possible through the assessment of a normalized, dimensionless quantity named $\phi_S$:

$$\phi_S = \vartheta_x^{-1}\int_{\mathbb{R}^3}d\mathbf{r}\sqrt{n_d(\mathbf{r})n_a(\mathbf{r})} \in [0;1]; \ \vartheta_x = \frac{1}{2}\sum_{q=d,\,a}\int_{\mathbb{R}^3}d\mathbf{r}\,n_q(\mathbf{r}) \tag{11}$$

where $\vartheta_x$ is a normalization factor (the integral of detachment/attachment density over all the space). Obviously, a long-range charge transfer means a low hole/particle overlap and will correspond to a low value for $\phi_S$. Conversely, a local transition will be characterized by a higher $\phi_S$ value. This is clearly illustrated in **Figure 1** where the $\phi_S$ value drops from 0.77 to 0.17 when going from an electronic transition exhibiting a large hole/particle overlap to a long-range charge transfer. These two cases are used solely to illustrate the potentiality of the $\phi_S$ quantum metric to assess the locality of a charge transfer. The computation of $\phi_S$ is schematically pictured in the top of **Figure 2**.

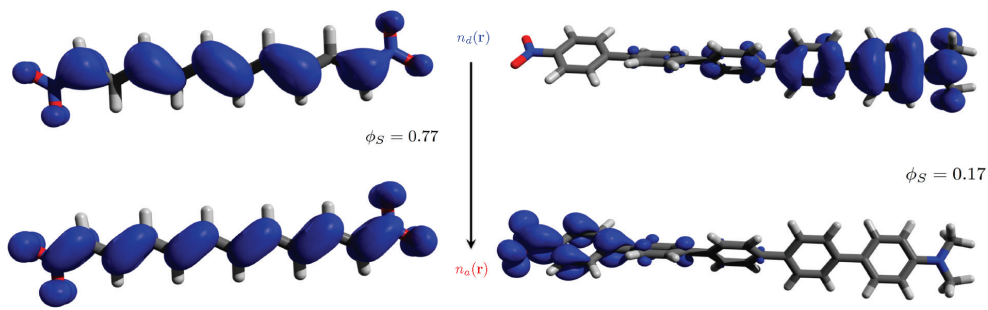

**Figure 1.** Illustration of a local (left) and long-range (right) transition using detachment/attachment densities and the $\phi_S$ index.

It has also been demonstrated that $\phi_S$ can be used for performing a diagnosis on the exchange-correlation functional used for computing the transition energy within the framework of TDDFT [3].

An additional quantitative strategy consists in computing the charge effectively displaced during the transition. The difference between the hole/particle and the effectively displaced charge density is illustrated in **Figure 2**: since there can be some overlap between the hole and the particle densities, the global outcome (the "bilan") of the transition in terms of charge displacement is not the detachment and attachment but the negative and positive contributions to the difference density, which can be obtained by taking the difference between the attachment and detachment charge densities at every point of space. Indeed, from

$$\mathbf{m} = \mathbf{k}_+ - \mathbf{k}_- \Rightarrow \gamma^\Delta = \mathbf{Mm}\mathbf{M}^\dagger = \mathbf{Mk}_+\mathbf{M}^\dagger - \mathbf{Mk}_-\mathbf{M}^\dagger = \gamma^a - \gamma^d \tag{12}$$

we can write

$$n_\Delta(\mathbf{r}) = n_a(\mathbf{r}) - n_d(\mathbf{r}), \tag{13}$$

and introduce the *actual* displacement charge density functions

$$n_\pm(\mathbf{r}) = \frac{1}{2}\left\{\sqrt{n_\Delta^2(\mathbf{r})} \pm n_\Delta(\mathbf{r})\right\} \tag{14}$$

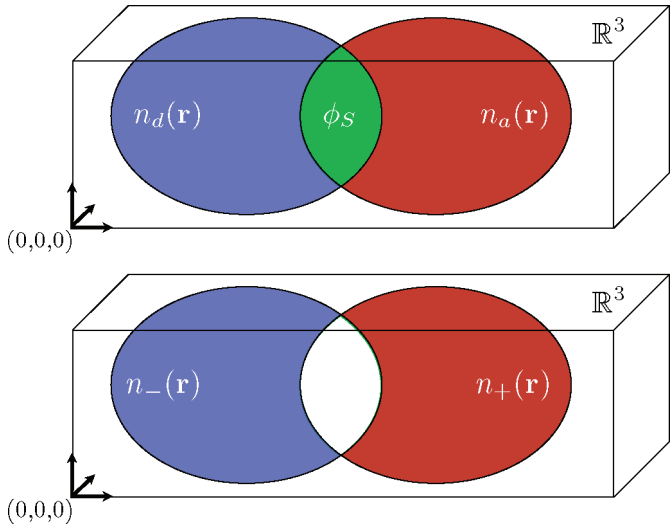

**Figure 2.** Illustration of the complementarity between $\phi_S$ and $\bar{\varphi}$.

so the splitting operation is performed based on the sign of function entries in the three dimensions of space instead of transition occupation numbers. From this separation we can compute the normalized displaced charge:

$$\frac{\vartheta_x^{-1}}{2} \sum_{s=+,\,-} \int_{\mathbb{R}^3} d\mathbf{r} \ n_s(\mathbf{r}) = \tilde{\varphi} \in [0;1]. \tag{15}$$

Obviously, splitting the transition occupation numbers and computing the detachment/attachment overlap are complementary to the integration of the negative and positive contributions to the difference density function: the $\phi_S$ descriptor provides an information related to the locality of the charge transfer, while the $\tilde{\varphi}$ metric relates the amount of charge transferred during the transition.

These two complementary approaches have been associated into a final, general quantum metric of charge transfer:

$$\psi = 2\pi^{-1} \underbrace{\arctan\left(\frac{\phi_S}{\tilde{\varphi}}\right)}_{\theta_S} = \frac{2\theta_S}{\pi} \in [0;1[ \tag{16}$$

which, as it was the case for $\phi_S$ and $\tilde{\varphi}$, is normalized and dimensionless. The $\psi$ metric can be interpreted as the normalized angle resulting from the joint projection of $\phi_S$ and $\tilde{\varphi}$ in a complex plane ($\tilde{\varphi}$ being along the real axis and $\phi_S$ the imaginary one). Such projection is characterized by a $\theta_S$ angle with the real axis (see Ref. [5]), taking values ranging from 0 to $\pi/2$. Therefore, the $2\pi^{-1}$ factor in Eq. (16) is there to ensure that $\psi$ is normalized. Note that there exists multiple ways to derive the three quantum metrics exposed in this paragraph, as mentioned in Ref. [5].

**Figure 3** represents the $\psi$ projection for a series of dyes. These chromophores are constituted by an electron-donor fragment conjugated to an acceptor moiety through a molecular bridge with a variable size (i.e., a variable number of subunits).

We see that when the first excited state of these dyes is computed using TDDFT with the hybrid PBE0 exchange-correlation functional [46, 47] and a triple-zeta split-valence Gaussian basis set with diffuse and polarization functions on every atom [48], increasing the number of bridge subunits leads to a net decrease in the $\psi$ projection angle. It is therefore very clear from **Figure 3** that increasing the length of the bridge for this family of dyes leads to an increase of the charge transfer character of the first transition, when computed at the above-mentioned level of theory.

The following paragraph details another known strategy providing a straightforward qualitative analysis of the charge transfer topology, based on another type of density matrix: the transition density matrix.

## 2.4. Transition density matrix and natural transition orbitals

In the following section we will be interested in the determination of the exciton wave function and its use for providing the most compact representation of an electronic transition. More particularly, this paragraph exposes how we can find an alternative basis to the canonical one and reduce the picture of the transition to one couple of hole/particle wave functions. The following formalism is applied to the case of quantum excited states that can be written as a linear combination of singly excited Slater determinants, constructed from the single-reference wave function ($\psi_0$) where the spinorbital $\varphi_i$ from the occupied canonical subspace has been replaced by the $\varphi_a$ spinorbital belonging to the virtual canonical subspace. In these conditions, the $x$th excited electronic state writes

$$|\psi_x\rangle = \sum_{i=1}^{N} \sum_{a=N+1}^{L} z_x^{-1/2} (\gamma^{0x})_{ia} |\psi_i^a\rangle; \ |\psi_i^a\rangle = \hat{a}^\dagger \hat{i} |\psi_0\rangle \qquad (17)$$

where again we introduced the annihilation $\hat{i}$ and creation $\hat{a}^\dagger$ operators from the second quantization, so we actually see that $|\psi_i^a\rangle$ is obtained by annihilating the electron in the $i$th spinorbital from the ground state wave function and creating an electron in the $a$th one. In Eq. (17),

$$z_x = \mathrm{tr}(\gamma^{0x} \gamma^{0x\dagger}) = \mathrm{tr}(\gamma^{0x\dagger} \gamma^{0x}) \qquad (18)$$

is a normalization factor and $(\gamma^{0x})_{ia}$ is a transition density matrix element for the $0 \rightarrow x$ state transition. Transition density matrix elements can be extracted from the exciton wave function:

$$\tilde{\gamma}^{0x}(\mathbf{r}_1, \mathbf{r}'_1) = N \sum_{\sigma_1 = \alpha, \beta} \int d\mathbf{x}_2 \dots \int d\mathbf{x}_N \, \psi_0(\mathbf{r}_1, \dots, \mathbf{x}_N) \psi_x^*(\mathbf{r}'_1, \dots, \mathbf{x}_N)$$

$$= z_x^{-1/2} \sum_{i=1}^{N} \sum_{a=N+1}^{L} \varphi_i(\mathbf{r}_1) \left(\boldsymbol{\gamma}^{0x}\right)_{ia} \varphi_a^*(\mathbf{r}'_1), \tag{19}$$

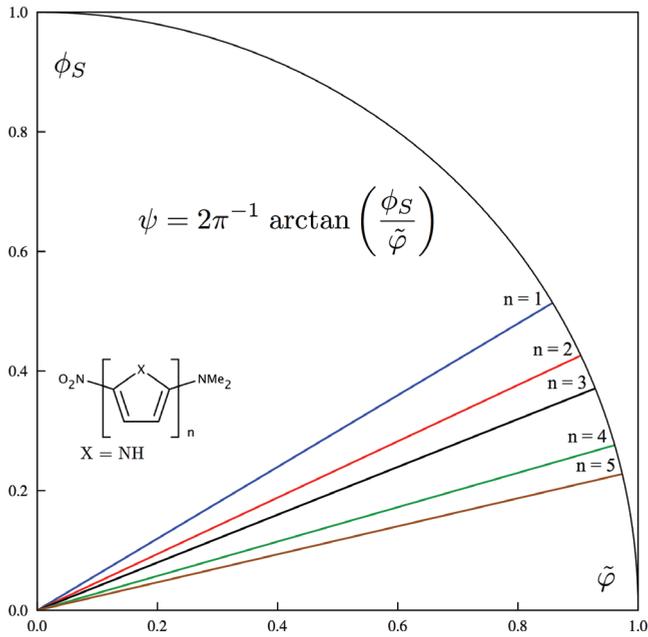

**Figure 3.** Illustration of the evolution of the $\psi$ index value for the first excited state of a series of push-pull dyes, computed at the PBE0/6-311++G(2d,p)//PBE0/6-311G(d,p) level of theory in vacuum.

That is, the so-called transition density matrix kernel locating the hole ($\mathbf{r}_1$) in the ground state and the particle ($\mathbf{r}'_1$) in the excited state. Similarly to the one-particle reduced density matrix in Eq. (4), the transition density matrix elements write

$$z_x^{-1/2}\left(\boldsymbol{\gamma}^{0x}\right)_{ia} = \left\langle \psi_0 | \hat{i}^\dagger \hat{a} | \psi_x \right\rangle = \sum_{j=1}^{N} \sum_{b=N+1}^{L} z_x^{-1/2} \left(\boldsymbol{\gamma}^{0x}\right)_{jb} \left\langle \psi_0 | \hat{i}^\dagger \hat{a} | \psi_j^b \right\rangle$$

$$= \sum_{j=1}^{N} \sum_{b=N+1}^{L} z_x^{-1/2} \left(\boldsymbol{\gamma}^{0x}\right)_{jb} \overbrace{\left\langle \psi_i^a | \psi_j^b \right\rangle}^{\delta_{ij}\delta_{ab}} = z_x^{-1/2} \left(\boldsymbol{\gamma}^{0x}\right)_{ia}. \tag{20}$$

Note that we conventionally set the $i, j$ and $a, b$ indices to match spinorbitals, respectively, belonging exclusively to the occupied and virtual canonical subspaces, while $r$ and $s$ indices have no restricted attribution to a given subspace. Similarly to the quantum state electron density function, one can deduce the expression of the one-particle transition density from the transition density matrix kernel:

$$n^{0x}(\mathbf{r}_1) = z_x^{-1/2} \sum_{i=1}^{N} \sum_{a=N+1}^{L} \varphi_i(\mathbf{r}_1) \left(\boldsymbol{\gamma}^{0x}\right)_{ia} \varphi_a^*(\mathbf{r}_1)$$

$$\Rightarrow \int_{\mathbb{R}^3} d\mathbf{r}_1 \; n^{0x}(\mathbf{r}_1) = z_x^{-1/2} \sum_{i=1}^{N} \sum_{a=N+1}^{L} \left(\boldsymbol{\gamma}^{0x}\right)_{ia} \underbrace{\left\langle \varphi_a | \varphi_i \right\rangle}_{\delta_{ia}} = 0 \tag{21}$$

where the $\delta_{ia}$ Kronecker delta is systematically vanishing since $\varphi_i$ and $\varphi_a$ spinorbitals never belong to the same subspace. Here again, we will take advantage of the possibility to use finite mathematical objects such as matrices and perform a reduction of the one-particle transition density matrix size: since we know that $i$ and $a$ indices are restricted to occupied and virtual subspaces, we can introduce the normalized transition density matrix $\mathbf{T}$ (that we will call transition density matrix in the following):

$$z_x^{-1/2}\left(\boldsymbol{\gamma}^{0x}\right)_{ia} \leftrightarrow (\mathbf{T})_{ic} \quad (c = a - N) \tag{22}$$

so the connection between the two matrices is trivial:

$$z_x^{-1/2} \boldsymbol{\gamma}^{0x} = \begin{pmatrix} 0_{N \times N} & \mathbf{T} \\ 0_{(L-N) \times N} & 0_{(L-N) \times (L-N)} \end{pmatrix}; \; z_x^{-1/2} \boldsymbol{\gamma}^{0x} \in \mathbb{R}^{L \times L} \leftrightarrow \mathbf{T} \in \mathbb{R}^{N \times (L-N)} \tag{23}$$

where $0_{k \times l}$ refers to the zero matrix with $k \times l$ dimensions. For the sake of simplicity, we will use $0_o$ and $0_v$ for the occupied $\times$ occupied and virtual $\times$ virtual zero blocks and $0_{o \times v}$ and $0_{v \times o}$ for the out-diagonal blocks.

We will now focus on $\mathbf{T}$. This matrix contains the information related to the transition we seek, and similarly to the difference density matrix, we will extract this information by diagonalizing $\mathbf{T}$. However, since $\mathbf{T}$ is not square but rectangular (we rarely have the same number of occupied and virtual orbitals), the diagonalization process is named singular value decomposition (SVD) [44] and takes the form

$$\exists \mathbf{O}, \mathbf{V} \mid \mathbf{O}^\dagger \mathbf{T} \mathbf{V} = \boldsymbol{\lambda}. \tag{24}$$

The diagonal $\boldsymbol{\lambda}$ entries are called the singular values of $\mathbf{T}$. Due to the dimensions of $\boldsymbol{\lambda}$, the number of singular values is equal to the dimensions of the lowest subspace (i.e., $N$ or $L - N$). Most often, the number of virtual orbitals is larger than the number of occupied orbitals. Therefore, from now on we will assume that $N < L - N$.

While from the diagonalization of $\gamma^A$ we could build detachment/attachment densities, here we will use the left and right eigenvectors of $\mathbf{T}$ for rotating the occupied and virtual canonical subspaces into the so-called occupied/virtual natural transition orbital (NTO) spaces:

$$\varphi_i^o(\mathbf{r}) = \sum_{j=1}^{N} (\mathbf{O})_{ji} \varphi_j(\mathbf{r}) \overset{(\boldsymbol{\lambda})_{ii}}{\longleftrightarrow} \varphi_i^v(\mathbf{r}) = \sum_{j=1}^{L-N} (\mathbf{V})_{ji} \varphi_{N+j}(\mathbf{r}), \tag{25}$$

where $i$ ranges from 1 to $N$. We have built $N$ couples of occupied/virtual NTOs, each couple being characterized by the corresponding singular value $(\boldsymbol{\lambda})_{ii}$. The great advantage of performing an SVD on $\mathbf{T}$ is that in most of the cases, only one singular value is predominant, which means that we can condensate all the physics of an electronic transition into one couple of occupied/virtual NTOs, as represented in **Figure 4**.

We can conclude that, similarly to the usual quantum state natural orbitals which constitute the basis in which the quantum state density matrix is diagonal, the NTOs provide the most compact representation of the electronic transition and can be used to rewrite the expression of the electronic excited state and the transition density matrix kernel (the exciton wave function):

$$|\psi_x\rangle = \sum_{i=1}^{N} (\boldsymbol{\lambda})_{ii} |\psi_{o,i}^{v,i}\rangle = \sum_{i=1}^{N} (\boldsymbol{\lambda})_{ii} \hat{q}_i^{v\dagger} \hat{q}_i^o |\psi_0\rangle; \; \tilde{\gamma}^{0x}(\mathbf{r}_1, \mathbf{r}_1') = \sum_{i=1}^{N} (\boldsymbol{\lambda})_{ii} \varphi_i^o(\mathbf{r}_1) \varphi_i^{v*}(\mathbf{r}_1') \tag{26}$$

where this time the creation/annihilation operators are bearing the "$o$" and "$v$" superscripts, reminding that we are annihilating an electron in the $i$th occupied ($o$) NTO and creating one electron in the $i$th virtual ($v$) NTO. Since we know that usually one singular value is predominant, we can clearly identify the hole and particle wave functions and state, upon light absorption, from where the electron goes and where it arrives.

Multiplying $\mathbf{T}$ by its own transpose and vice versa leads to two square matrices with interesting properties:

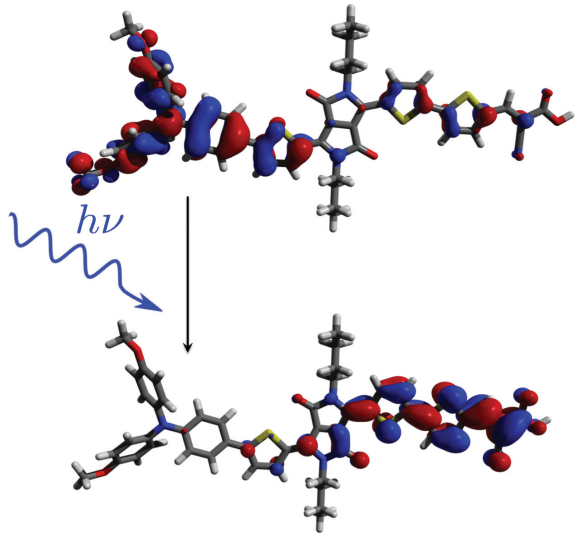

**Figure 4.** Illustration of the hole (top) and particle (bottom) wave functions, that is, the predominant couple of occupied (top) and virtual (bottom) NTOs for a random push-pull chromophore experiencing a photoinduced charge transfer.

$$\mathbf{T}\mathbf{T}^{\dagger} \in \mathbb{R}^{N \times N}; \ \mathbf{T}^{\dagger}\mathbf{T} \in \mathbb{R}^{(L-N) \times (L-N)}. \tag{27}$$

Due to their structure, these two new matrices share the same eigenvectors than $\mathbf{T}$

$$\mathbf{O}^{\dagger}\mathbf{T}\mathbf{T}^{\dagger}\mathbf{O} = \lambda_{o}^{2}; \ \mathbf{V}^{\dagger}\mathbf{T}^{\dagger}\mathbf{T}\mathbf{V} = \lambda_{v}^{2} \tag{28}$$

with, considering $N < L - N$, the following rules for their eigenvalues:

$$(\lambda)_{ii}^{2} = (\lambda_{o}^{2})_{ii} = (\lambda_{v}^{2})_{ii} \ \forall i \leq N; \quad \lambda_{v}^{2} = \lambda_{o}^{2} \oplus 0_{v}. \tag{29}$$

These rules can be demonstrated by developing the product of $\lambda$ with its own transpose:

$$\lambda\lambda^{\dagger} = \mathbf{O}^{\dagger}\mathbf{T}\underbrace{\mathbf{V}\mathbf{V}^{\dagger}}_{I_{v}} \mathbf{T}^{\dagger}\mathbf{O} = \mathbf{O}^{\dagger}\mathbf{T}\mathbf{T}^{\dagger}\mathbf{O} \tag{30}$$

where $I_{v}$ is the $(L-N) \times (L-N)$ identity matrix. Due to the dimensions of $\lambda$ and its diagonal structure, we can write

$$\lambda \in \mathbb{R}^{N \times (L-N)} \ \Rightarrow \ \lambda\lambda^{\dagger} \in \mathbb{R}^{N \times N} \quad ; \quad (\lambda)_{ij} = 0 \ \forall i \neq j \ \Rightarrow \ \lambda\lambda^{\dagger} = \lambda_{o}^{2}. \tag{31}$$

Similarly, we have for $\lambda^{\dagger}\lambda$

$$\lambda^{\dagger}\lambda = \mathbf{V}^{\dagger}\mathbf{T}^{\dagger}\underbrace{\mathbf{O}\mathbf{O}^{\dagger}}_{I_{o}} \mathbf{T}\mathbf{V} = \mathbf{V}^{\dagger}\mathbf{T}^{\dagger}\mathbf{T}\mathbf{V} \tag{32}$$

and

$$\lambda \in \mathbb{R}^{N \times (L-N)} \;\Rightarrow\; \lambda^\dagger \lambda \in \mathbb{R}^{(L-N) \times (L-N)} \quad; \quad (\lambda)_{ij} = 0 \;\; \forall i \neq j \;\Rightarrow\; \lambda^\dagger \lambda = \lambda_v^2. \tag{33}$$

Multiplying Eq. (28) by the left by $\mathbf{T}^\dagger \mathbf{O}$ or $\mathbf{T}\mathbf{V}$ leads to two new eigenvalue problems:

$$\mathbf{T}^\dagger \mathbf{O}(\mathbf{O}^\dagger \mathbf{T}\mathbf{T}^\dagger \mathbf{O} = \lambda_o^2) \;\Leftrightarrow\; \mathbf{T}^\dagger \mathbf{T}\underbrace{\mathbf{T}^\dagger \mathbf{O}}_{\mathbf{V}_o} = \mathbf{T}^\dagger \mathbf{O}\lambda_o^2 \quad; \quad \mathbf{T}\mathbf{V}(\mathbf{V}^\dagger \mathbf{T}^\dagger \mathbf{T}\mathbf{V} \Leftrightarrow \lambda_v^2) = \mathbf{T}\mathbf{T}^\dagger \underbrace{\mathbf{T}\mathbf{V}}_{\mathbf{O}_v} = \mathbf{T}\mathbf{V}\lambda_v^2 \tag{34}$$

where $\mathbf{V}_o \in \mathbb{R}^{(L-N) \times N}$ contains the $N$ eigenvectors of $\mathbf{T}^\dagger \mathbf{T}$ with a nonvanishing eigenvalue (i.e., the $N$ first columns of $\mathbf{V}$) and $\mathbf{O}_v \in \mathbb{R}^{N \times (L-N)}$ is the juxtaposition of $\mathbf{O}$ and $L - 2N$ zero columns. The results in Eq. (34) prove that the eigenvectors of each of the two matrices in Eq. (27) can be found from the eigenvectors of the other one and that both matrices share the same nonvanishing eigenvalues, as mentioned in Eq. (29).

# 3. Bridging the detachment/attachment and NTO paradigms

We now have two general strategies for qualitatively studying the topology of the light-induced electronic cloud polarization, and the locality of this electronic structure reorganization can be quantified. This section is devoted to single-reference excited states calculation methods that express the electronic excited state as a linear combination of singly excited Slater determinants and brings the rigorous demonstration that in such case, the three quantum metrics we previously designed can be formally equivalently derived from the difference density matrix or the transition density matrix. This result is the corollary to a theorem stating that the occupied/virtual NTOs are nothing but the eigenvectors of the detachment/attachment density matrices.

## 3.1. Expression of the quantum state density matrices in the canonical space

In this paragraph we elucidate the structure of the difference density matrix by developing the full expression of the excited state density matrix in the canonical space.

**Lemma III.1** *The difference density matrix is the direct sum of* $-\mathbf{T}\mathbf{T}^\dagger$ *and* $\mathbf{T}^\dagger \mathbf{T}$.

*Proof.* We start by writing the expression of the ground state density matrix: from Eq. (4) it follows that for an $N$–electron single-determinant ground state wave function,

$$\forall (r,s) \mid r \leq N \text{ and } s \leq N, \;\; (\gamma)_{rs} = \delta_{rs} \quad; \quad \forall (r,s) \mid r > N \text{ and/or } s > N, \;\; (\gamma)_{rs} = 0. \tag{35}$$

It follows that the ground state one-particle density matrix in the canonical space writes

$$\gamma^0 = I_o \oplus 0_v. \tag{36}$$

If now we rewrite the electronic excited state $|\psi_x\rangle$ from Eq. (17) using the normalized transition density matrix elements, we have

$$|\psi_x\rangle = \sum_{i=1}^{N} \sum_{a=N+1}^{L} (\mathbf{T})_{ic} |\psi_i^a\rangle \quad (c = a - N). \tag{37}$$

From now on we will operate a systematic index shift between matrix elements and virtual orbitals implied in the singly excited Slater determinants. Since the excited state wave function is normalized, we can write

$$1 = \langle\psi_x|\psi_x\rangle = \sum_{i,j=1a,}^{N} \sum_{b=N+1}^{L} (\mathbf{T})_{jd}^{*} (\mathbf{T})_{ic} \overbrace{\langle\psi_j^b|\psi_i^a\rangle}^{\delta_{ij}\delta_{ab}} \qquad (d = b - N) \\ = \sum_{i=1}^{N} \sum_{a=N+1}^{L} (\mathbf{T})_{ic}^{*} (\mathbf{T})_{ic} = \sum_{i=1}^{N} \sum_{a=N+1}^{L} (\mathbf{T})_{ic} (\mathbf{T}^{\dagger})_{ci} = \begin{cases} \text{tr}(\mathbf{T}\mathbf{T}^{\dagger}) \\ \text{tr}(\mathbf{T}^{\dagger}\mathbf{T}) \end{cases} \tag{38}$$

and, since the trace of a matrix is an unitary invariant,

$$\text{tr}(\boldsymbol{\lambda}\boldsymbol{\lambda}^{\dagger}) = \text{tr}(\boldsymbol{\lambda}^{\dagger}\boldsymbol{\lambda}) = 1. \tag{39}$$

Using the second quantization, we might rewrite $|\psi_x\rangle$

$$|\psi_x\rangle = \sum_{i=1}^{N} \sum_{a=N+1}^{L} (\mathbf{T})_{ic} \ \hat{a}^{\dagger}\hat{i}|\psi_0\rangle \tag{40}$$

and the $r \times s$ density matrix element for the $x$th excited state writes

$$(\boldsymbol{\gamma}^x)_{rs} = \left\langle\psi_x|\hat{r}^{\dagger}\hat{s}|\psi_x\right\rangle = \sum_{i,j=1a,}^{N} \sum_{b=N+1}^{L} (\mathbf{T})_{jd}^{*} (\mathbf{T})_{ic} \left\langle\psi_j^b|\hat{r}^{\dagger}\hat{s}|\psi_i^a\right\rangle \\ = \sum_{i,j=1a,}^{N} \sum_{b=N+1}^{L} (\mathbf{T})_{jd}^{*} (\mathbf{T})_{ic} \left\langle\psi_0|\hat{j}^{\dagger}\hat{b}\hat{r}^{\dagger}\hat{s}\hat{a}^{\dagger}\hat{i}|\psi_0\right\rangle. \tag{41}$$

We will now apply Wick's theorem to the expression of the excited state density matrix written using our fermionic second quantization operators. According to this theorem, one can rewrite Eq. (41) as a combination of products of expectation values of couples of the second quantization operators implied in the expression of $\boldsymbol{\gamma}^x$. Since we are working with fermionic operators, a phase is assigned to each term of this sum with the form $(-1)^{q_l}$ where $l$ corresponds to the position of the term in the sum. Note that a number is also assigned to the position of each fermionic operator both in the original expression of $\boldsymbol{\gamma}^x$ and after expanding it into a sum of terms. **Figure 5** illustrates the case of $\boldsymbol{\gamma}^x$, which can be decomposed into a sum of three nonvanishing terms. The central part of the figure shows how each term is constructed by associating a creation to an annihilation operator. Note that other operator pairings are possible, but their expectation value is vanishing due to the fact that the associated operators do not belong to the same subspace (occupied or virtual). The right part of **Figure 5** shows how the label sequence of the operators has been rearranged for each term.

$$\langle\psi_0|\hat{j}^\dagger\hat{b}\hat{r}^\dagger\hat{s}\hat{a}^\dagger\hat{i}|\psi_0\rangle_{\,1\,2\,3\,4\,5\,6}=\left\{\begin{array}{c}\langle\psi_0|\hat{j}^\dagger\hat{b}\hat{r}^\dagger\hat{s}\hat{a}^\dagger\hat{i}|\psi_0\rangle_{\,1\,2\,3\,4\,5\,6}\\[2em]\langle\psi_0|\hat{j}^\dagger\hat{b}\hat{r}^\dagger\hat{s}\hat{a}^\dagger\hat{i}|\psi_0\rangle_{\,1\,2\,3\,4\,5\,6}\\[2em]\langle\psi_0|\hat{j}^\dagger\hat{b}\hat{r}^\dagger\hat{s}\hat{a}^\dagger\hat{i}|\psi_0\rangle_{\,1\,2\,3\,4\,5\,6}\end{array}\right.$$

$$=\;\begin{array}{l}(-1)^{\varrho_1}\,\langle\psi_0|\hat{j}^\dagger\hat{i}|\psi_0\rangle_{\,1\,6}\,\langle\psi_0|\hat{b}\hat{a}^\dagger|\psi_0\rangle_{\,2\,5}\,\langle\psi_0|\hat{r}^\dagger\hat{s}|\psi_0\rangle_{\,3\,4}\\[1.5em]+(-1)^{\varrho_2}\,\langle\psi_0|\hat{j}^\dagger\hat{s}|\psi_0\rangle_{\,1\,4}\,\langle\psi_0|\hat{b}\hat{a}^\dagger|\psi_0\rangle_{\,2\,5}\,\langle\psi_0|\hat{r}^\dagger\hat{i}|\psi_0\rangle_{\,3\,6}\\[1.5em]+(-1)^{\varrho_3}\,\langle\psi_0|\hat{j}^\dagger\hat{i}|\psi_0\rangle_{\,1\,6}\,\langle\psi_0|\hat{s}\hat{a}^\dagger|\psi_0\rangle_{\,4\,5}\,\langle\psi_0|\hat{b}\hat{r}^\dagger|\psi_0\rangle_{\,2\,3}\end{array}$$

**Figure 5.** Wick's theorem applied to single-reference excited state density matrices.

Once the excited state density matrix is developed, one can write a bijection $f_l(x)=y$ between the original sequence of operators label (here 1, …, 6) and the one characterizing each term ($l=1,2,3$). The $\varrho_l$ value is then obtained by counting the number of pairs of projections satisfying

$$(x_1, x_2)\ |\ \{x_1 < x_2;\ f_l(x_1) > f_l(x_2)\} \tag{42}$$

in the bijection. For example, for the first term ($l=1$), the ($x_1=2$, $x_2=5$) pair satisfies this condition, because $f_1(2)=6>3=f_1(5)$. The evaluation of the phase to be assigned to the first term ($l=1$) reported in **Figure 5** is fully detailed in **Figure 6**. The deduction of the phase for $l=2$ and 3 is given in Appendix (**Figures 7** and **8**).

For each term in the developed expression of $\gamma^x$, six permutations of its factors are possible without affecting the phase, for the parity of $\varrho_l$ is guided only by the primary association of creation/annihilation operators characterizing the $l$th term. According to what precedes, we are now able to write the $r \times s$ excited state density matrix element:

$$\left(\gamma^x\right)_{rs} = \sum_{i,\,j=1}^{N}\sum_{a,\,b=N+1}^{L}(\mathbf{T})_{jd}^*\,(\mathbf{T})_{ic}(\mathscr{I}_1 - \mathscr{I}_2 + \mathscr{I}_3) \tag{43}$$

with

$$\mathscr{I}_1 = \overbrace{\left\langle\psi_0|\hat{j}^\dagger\hat{i}|\psi_0\right\rangle}^{\delta_{ij}}\underbrace{\left\langle\psi_0|\hat{b}\hat{a}^\dagger|\psi_0\right\rangle}_{\delta_{ab}}\overbrace{\left\langle\psi_0|\hat{r}^\dagger\hat{s}|\psi_0\right\rangle}^{\delta_{rs}n_r} \tag{44}$$

where $n_r$ is the occupation number of spinorbital $r$ (see Eq. (35) for more details). For $\mathscr{I}_2$ we have

$$\mathscr{I}_2 = \overbrace{\left\langle \psi_0 | \hat{j}^\dagger \hat{s} | \psi_0 \right\rangle}^{\delta_{js}} \underbrace{\left\langle \psi_0 | \hat{b} \hat{a}^\dagger | \psi_0 \right\rangle}_{\delta_{ab}} \overbrace{\left\langle \psi_0 | \hat{r}^\dagger \hat{i} | \psi_0 \right\rangle}^{\delta_{ri}} \tag{45}$$

and for $\mathscr{I}_3$,

$$\mathscr{I}_3 = \overbrace{\left\langle \psi_0 | \hat{j}^\dagger \hat{i} | \psi_0 \right\rangle}^{\delta_{ij}} \underbrace{\left\langle \psi_0 | \hat{s} \hat{a}^\dagger | \psi_0 \right\rangle}_{\delta_{sa}} \overbrace{\left\langle \psi_0 | \hat{b} \hat{r}^\dagger | \psi_0 \right\rangle}^{\delta_{br}}. \tag{46}$$

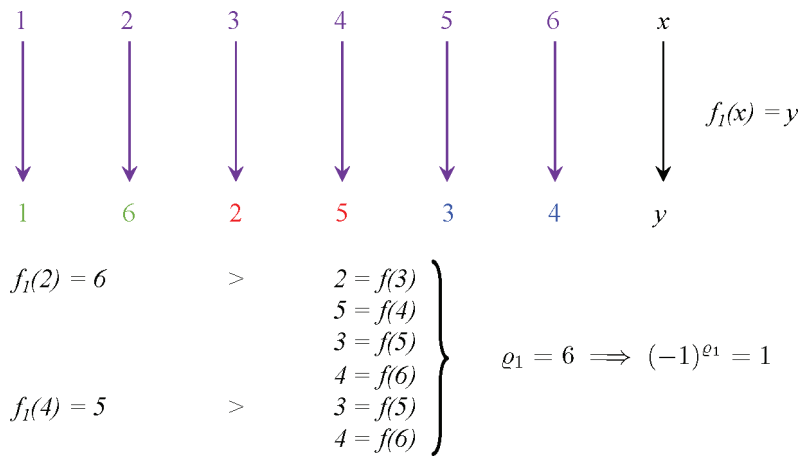

**Figure 6.** Illustration of the evaluation of $\varrho_1$.

Note that since $i$ and $j$ are corresponding to occupied spinorbitals, writing $\delta_{js}$ is equivalent to writing $\delta_{js}n_s$ and is not vanishing only when $\varphi_s$ belongs to the occupied subspace. This is also the case for $\delta_{ri}$. On the other hand, since $\varphi_a$ and $\varphi_b$ belong to the virtual subspace, writing $\delta_{sa}$ is equivalent to writing $\delta_{sa}(1 - n_s)$ and is not vanishing only when $s$ is superior to $N$. Note also that writing $\delta_{ab}$ when dealing with spinorbitals corresponds to $\delta_{cd}$ when working with matrix elements (see Eqs. (37) and (38)). Therefore, $(\gamma^x)_{rs}$ now writes

$$
\begin{aligned}
(\boldsymbol{\gamma}^x)_{rs} = \delta_{rs} n_r &\overbrace{\left[ \sum_{i,\,j=1\,a,\,}^{N} \sum_{b=N+1}^{L} (\mathbf{T})_{jd}^* (\mathbf{T})_{ic} \delta_{ij} \delta_{ab} \right]}^{\mathrm{tr}(\mathbf{TT}^\dagger) = \mathrm{tr}(\mathbf{T}^\dagger \mathbf{T}) = 1} \\
&- \sum_{i,\,j=1\,a,\,}^{N} \sum_{b=N+1}^{L} (\mathbf{T})_{jd}^* (\mathbf{T})_{ic} \delta_{ab} \delta_{js} n_s \delta_{ri} n_r \\
&+ \sum_{i,\,j=1\,a,\,}^{N} \sum_{b=N+1}^{L} (\mathbf{T})_{jd}^* (\mathbf{T})_{ic} \delta_{ij} \delta_{br} (1 - n_r) \delta_{sa} (1 - n_s),
\end{aligned}
\tag{47}
$$

that is,

$$\left(\boldsymbol{\gamma}^x\right)_{rs} = \delta_{rs} n_r - \left(\mathbf{TT}^\dagger\right)_{ij} \delta_{js} n_s \delta_{ri} n_r + \left(\mathbf{T}^\dagger\mathbf{T}\right)_{dc} (1 - n_r) \delta_{d(r-N)} (1 - n_s) \delta_{c(s-N)}. \tag{48}$$

We see that the first and second terms belong to the occupied × occupied block, while the third term belongs to the virtual × virtual one. According to this, the excited state density matrix in the canonical space finally writes

$$\boldsymbol{\gamma}^x = \left(I_o - \mathbf{TT}^\dagger\right) \oplus \mathbf{T}^\dagger\mathbf{T}. \tag{49}$$

Subtracting the ground state density matrix taken from Eq. (36) to $\boldsymbol{\gamma}^x$ gives $\boldsymbol{\gamma}^\Delta$

$$\boldsymbol{\gamma}^\Delta = -\mathbf{TT}^\dagger \oplus \mathbf{T}^\dagger\mathbf{T}. \quad\blacksquare \tag{50}$$

Since $\mathbf{TT}^\dagger$ and $\mathbf{T}^\dagger\mathbf{T}$ have positive eigenvalues (i.e., they are positive definite), we deduce

$$(\mathbf{m})_{ii} \leq 0 \;\; \forall i \leq N; \;\; (\mathbf{m})_{aa} \geq 0 \;\; \forall a > N. \tag{51}$$

Therefore, we must have that

$$\mathbf{TT}^\dagger \oplus 0_v = \boldsymbol{\gamma}^d \quad ; \quad 0_o \oplus \mathbf{T}^\dagger\mathbf{T} = \boldsymbol{\gamma}^a \tag{52}$$

which obviously leads to

$$\boldsymbol{\gamma}^\Delta = \boldsymbol{\gamma}^a - \boldsymbol{\gamma}^d. \tag{53}$$

This last statement is in agreement with (12). Note that

$$\sum_{r=1}^{N} \left(\mathbf{TT}^\dagger\right)_{rr} = \int_{\mathbb{R}^3} d\mathbf{r} \; n_d(\mathbf{r}) = \vartheta_x = \int_{\mathbb{R}^3} d\mathbf{r} \; n_a(\mathbf{r}) = \sum_{s=1}^{L-N} \left(\mathbf{T}^\dagger\mathbf{T}\right)_{ss}. \tag{54}$$

It follows that $\vartheta_x = 1$.

### 3.2. Detachment/attachment density matrix eigenvectors

This paragraph aims at demonstrating the connection between the NTOs and detachment/attachment paradigms by using the structure of the difference density matrix.

**Theorem III.1** *NTOs are the eigenvectors of the detachment/attachment density matrices.*

*Proof.* We know from Lemma III.1 that

$$\boldsymbol{\gamma}^\Delta = -\mathbf{TT}^\dagger \oplus \mathbf{T}^\dagger\mathbf{T} = \overbrace{\begin{pmatrix} 0_o & 0_{o\times v} \\ 0_{v\times o} & \mathbf{T}^\dagger\mathbf{T} \end{pmatrix}}^{\boldsymbol{\gamma}^a} - \underbrace{\begin{pmatrix} \mathbf{TT}^\dagger & 0_{o\times v} \\ 0_{v\times o} & 0_v \end{pmatrix}}_{\boldsymbol{\gamma}^d}; \quad \exists \mathbf{M} \mid \mathbf{M}^\dagger \boldsymbol{\gamma}^\Delta \mathbf{M} = \mathbf{m}. \tag{55}$$

Since $\mathbf{TT}^\dagger$ and $\mathbf{T}^\dagger\mathbf{T}$ are positive definite, we deduce that the only negative eigenvalues of $\gamma^\Delta$ belong to the occupied $\times$ occupied block, while the positive ones belong to the virtual $\times$ virtual block. Since we know how to obtain the eigenvalues of $\mathbf{TT}^\dagger$ and $\mathbf{T}^\dagger\mathbf{T}$ thanks to Eq. (28), we know that the matrix $\mathbf{M}$ diagonalizing the difference density matrix must be the direct sum of $\mathbf{O}$ and $\mathbf{V}$:

$$\mathbf{m} = -\lambda_o^2 \oplus \lambda_v^2 \quad ; \quad \mathbf{M} = \mathbf{O} \oplus \mathbf{V}. \quad \blacksquare \tag{56}$$

According to Eq. (52), we deduce that the eigenvectors of the detachment/attachment density matrices are nothing but the occupied/virtual natural transition orbitals: the $\mathbf{M}^{d,a}$ matrices diagonalizing $\gamma^{d,a}$ are

$$\mathbf{M}^d = \mathbf{O} \oplus 0_v \;\Rightarrow\; \mathbf{M}^{d\dagger}\gamma^d\mathbf{M}^d = \lambda_0^2 \oplus 0_v; \quad \mathbf{M}^a = 0_o \oplus \mathbf{V} \;\Rightarrow\; \mathbf{M}^{a\dagger}\gamma^a\mathbf{M}^a = 0_o \oplus \lambda_v^2. \tag{57}$$

### 3.3. Equivalence of the two paradigms through quantitative analysis

Finally, and since we demonstrated that there is a direct relationship between the NTOs and the detachment/attachment, we will use Lemma III.1 and Theorem III.1 to demonstrate that our quantitative analysis is equivalent when derived in the two paradigms when the ground state wave function is a single Slater determinant and the excited state is a normalized linear combination of singly excited Slater determinants.

**Corollary III.1** *The quantum descriptors derived from $\gamma^\Delta$ can be derived from $\mathbf{T}$'s eigenvectors and singular values.*

*Proof.* From Lemma III.1 and Theorem III.1, we can construct the following scheme:

$$\mathbf{O}^\dagger\mathbf{TV} = \lambda \rightarrow \{\lambda_o^2; \lambda_v^2\} \rightarrow \gamma^\Delta = -\mathbf{O}\lambda_o^2\mathbf{O}^\dagger \oplus \mathbf{V}\lambda_v^2\mathbf{V}^\dagger. \tag{58}$$

Following the structure of $\mathbf{m}$ deduced in Theorem III.1, we simply find $\mathbf{k}_\pm$

$$\mathbf{m} = -\lambda_o^2 \oplus \lambda_v^2 \Rightarrow \mathbf{k}_+ = 0_o \oplus \lambda_v^2 \quad ; \quad \mathbf{k}_- = \lambda_o^2 \oplus 0_v. \tag{59}$$

Backtransformation and few manipulations lead to

$$(\mathbf{O} \oplus \mathbf{V})\mathbf{k}_\pm(\mathbf{O}^\dagger \oplus \mathbf{V}^\dagger) = \gamma^{d,a} \rightarrow \{\phi_S, \tilde{\varphi}, \psi\}. \quad \blacksquare \tag{60}$$

The joint computation of the NTOs and detachment/attachment density matrices from a single SVD, as a preliminary to the quantum metrics assessment, can even be simplified as

$$\mathbf{O}^\dagger\mathbf{TV} = \lambda \rightarrow (\mathbf{O} \oplus \mathbf{V}) \begin{Bmatrix} \lambda\lambda^\dagger & \oplus & 0_v \\ 0_o & \oplus & \lambda^\dagger\lambda \end{Bmatrix} (\mathbf{O}^\dagger \oplus \mathbf{V}^\dagger) = \begin{Bmatrix} \gamma^d \\ \gamma^a \end{Bmatrix} \rightarrow \{\phi_S, \tilde{\varphi}, \psi\}. \tag{61}$$

Note finally that from Eq. (52) we see that the computation of the detachment/attachment density matrices (hence, the assessment of the topological metrics) can be performed without requiring any matrix diagonalization.

## 4. Conclusion

We rigorously detailed the theoretical background related to two methods allowing one to straightforwardly visualize how the absorption or emission of a photon impacts the electronic distribution of any complex molecular system. Based on one of these two methods, we showed that quantitative insights can be easily reached. Subsequently, we bridged the formalism of our two qualitative strategies in the case of single-reference excited states methods solely involving singly excited Slater determinants. Finally, it was demonstrated that in these cases any of the two qualitative methods can be used as a basis for deriving equivalent quantitative results. The totality of the features exposed in this book chapter is currently coded in the Nancy-Ex 2.0 [49] software suite and will be revisited, together with new strategies, in the TÆLES software [50] to be published soon.

## Acknowledgements

Prof. Xavier Assfeld and Drs. Antonio Monari, Mariachiara Pastore, Benjamin Lasorne, Matthieu Saubanère, and Felix Plasser are gratefully acknowledged for fruitful discussions on the topic. Profs. Istvan Mayer, Anatoliy Luzanov, Martin Head-Gordon, and Andreas Dreuw are also thanked for their extremely inspiring work.

## A. Derivation of the phase for $l = 2$ and $3$

**Figures 7** and **8** illustrate the evaluation process for the phase of terms 2 and 3 of Wick's expansion of the excited state density matrix elements in Eq. (41).

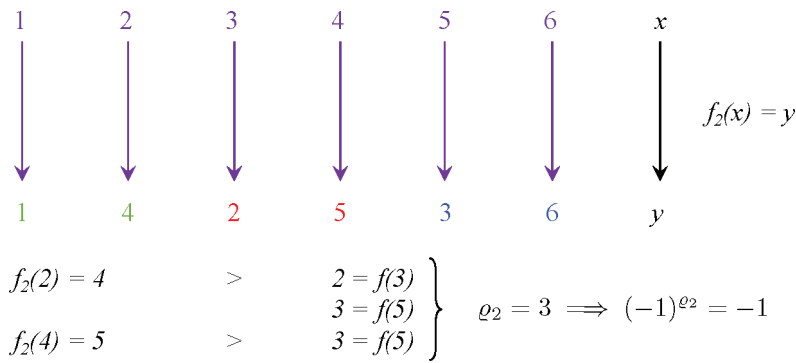

**Figure 7.** Illustration of the evaluation of $\varrho_2$.

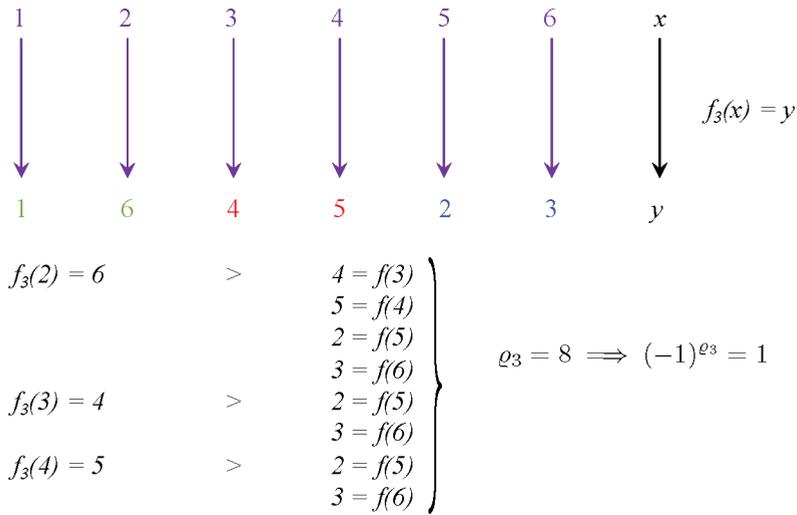

**Figure 8.** Illustration of the evaluation of $\varrho_3$.

## B. Derivation of the equations in the atomic orbitals space

Most of the time, the spinorbitals themselves are expressed in a basis (often called basis of atomic orbitals, basis of atomic functions, or more simply a basis set) of $K$ functions $\{\phi\}$. $K$ might be superior to $L$ when multiple spinorbitals in the atomic space are linearly dependent. The expression of spinorbitals in the atomic space is called linear combination of atomic orbitals (LCAO), and the pondering coefficients for a given spinorbital are stored in the column of a matrix, $\mathbf{C} \in \mathbb{R}^{K \times L}$, so that any spinorbital writes

$$\varphi_l(\mathbf{r}) = \sum_{\mu=1}^{K} (\mathbf{C})_{\mu l} \ \phi_\mu(\mathbf{r}). \tag{62}$$

Note that atomic orbitals are denoted by Greek letters for matrix elements. Since the spinorbitals correspond to columns in $\mathbf{C}$, we can split $\mathbf{C}$ into two matrices, $\widetilde{\mathbf{O}} \in \mathbb{R}^{K \times N}$ and $\widetilde{\mathbf{V}} \in \mathbb{R}^{K \times (L-N)}$, where the former contains the LCAO coefficients of the $N$ first spinorbitals (the occupied ones) and the latter contains the LCAO coefficients for the last $L - N$ spinorbitals (the virtual ones). This splitting operation will be used later.

The spatial overlap between two atomic functions is also stored into a matrix, $\mathbf{S}$, which has the following elements:

$$(\mathbf{S})_{\mu\nu} = \int_{\mathbb{R}^3} d\mathbf{r} \ \phi_\mu^*(\mathbf{r})\phi_\nu(\mathbf{r}). \tag{63}$$

According to the LCAO expansion, the one-particle reduced density matrix kernel from Eq. (1) can be written in the atomic space for defining the density matrix $\mathbf{P}$ in the atomic space

$$\tilde{\gamma}(\mathbf{r}_1,\mathbf{r}_1') = \sum_{r=1}^{L}\sum_{s=1}^{L}\sum_{\mu=1}^{K}\sum_{\nu=1}^{K}\phi_\mu(\mathbf{r}_1)(\mathbf{C})_{\mu r}(\boldsymbol{\gamma})_{rs}(\mathbf{C})_{\nu s}^*\phi_\nu^*(\mathbf{r}_1') = \sum_{\mu=1}^{K}\sum_{\nu=1}^{K}\phi_\mu(\mathbf{r}_1)\underbrace{\sum_{r=1}^{L}\sum_{s=1}^{L}(\mathbf{C})_{\mu r}(\boldsymbol{\gamma})_{rs}(\mathbf{C}^\dagger)_{s\nu}}_{(\mathbf{P})_{\mu\nu}}\phi_\nu^*(\mathbf{r}_1')$$

$$= \sum_{\mu=1}^{K}\sum_{\nu=1}^{K}(\mathbf{P})_{\mu\nu}\phi_\mu(\mathbf{r}_1)\phi_\nu^*(\mathbf{r}_1'). \tag{64}$$

In these conditions, the number of electrons is given by the trace of $\mathbf{PS}$. The central object for our investigations is now $\mathbf{P}$, so that in the atomic space, the difference density matrix writes

$$\boldsymbol{\Delta} = \mathbf{P}^x - \mathbf{P}^0 \Rightarrow \mathrm{tr}(\boldsymbol{\Delta}\mathbf{S}) = 0. \tag{65}$$

The difference density matrix in the atomic space can be diagonalized

$$\exists \mathbf{U} \mid \mathbf{U}^\dagger\boldsymbol{\Delta}\mathbf{U} = \boldsymbol{\delta}. \tag{66}$$

Note here that $\boldsymbol{\delta}$ is a diagonal matrix containing the $\boldsymbol{\Delta}$ eigenvalues and should not be confused with the Kronecker delta. The $\boldsymbol{\Delta}$ eigenvalues can be sorted according to their sign:

$$\boldsymbol{\sigma}_\pm = \frac{1}{2}\left(\sqrt{\boldsymbol{\delta}^2} \pm \boldsymbol{\delta}\right) \tag{67}$$

and the resulting diagonal matrices can be separately backtransformed to provide the so-called detachment ($\mathbf{D}$) and attachment ($\mathbf{A}$) density matrices and the corresponding charge densities:

$$\mathbf{U}\boldsymbol{\sigma}_-\mathbf{U}^\dagger = \mathbf{D} \xrightarrow{\mathbb{R}^3} n_d(\mathbf{r}) = \sum_{\mu=1}^{K}\sum_{\nu=1}^{K}(\mathbf{D})_{\mu\nu}\ \phi_\mu(\mathbf{r})\phi_\nu^*(\mathbf{r}) \quad ; \quad \mathbf{U}\boldsymbol{\sigma}_+\mathbf{U}^\dagger = \mathbf{A} \xrightarrow{\mathbb{R}^3} n_a(\mathbf{r}) = \sum_{\mu=1}^{K}\sum_{\nu=1}^{K}(\mathbf{A})_{\mu\nu}\ \phi_\mu(\mathbf{r})\phi_\nu^*(\mathbf{r}). \tag{68}$$

From the detachment and attachment charge densities, one can then compute $\phi_S$, $\tilde{\varphi}$, and $\psi$. Note that $\mathbf{D}$ and $\mathbf{A}$ should not be confused with "Donor" and "Acceptor" when dealing with push-pull dyes since, as we saw in **Figure 1**, the detachment or attachment densities are not strictly localized on fragments. Indeed, the detachment/attachment analysis is said to be systematic (or global), so is the quantitative analysis derived from it.

According to the structure of $\boldsymbol{\gamma}^\Delta$ derived in Lemma III.1, and the connection between density matrices in the canonical and atomic spaces (see Eq. (64)), we can write $\boldsymbol{\Delta}$ using $\mathbf{T}$:

$$\boldsymbol{\Delta} = \mathbf{C}\boldsymbol{\gamma}^\Delta\mathbf{C}^\dagger = \mathbf{C}\left(-\mathbf{T}\mathbf{T}^\dagger \oplus \mathbf{T}^\dagger\mathbf{T}\right)\mathbf{C}^\dagger \tag{69}$$

which reduces to

$$\boldsymbol{\Delta} = \tilde{\mathbf{V}}\left(\mathbf{T}^\dagger\mathbf{T}\right)\tilde{\mathbf{V}}^\dagger - \tilde{\mathbf{O}}\left(\mathbf{T}\mathbf{T}^\dagger\right)\tilde{\mathbf{O}}^\dagger. \tag{70}$$

This means that from the transition density matrix one can easily reconstruct the difference density matrix in the atomic space, diagonalize it, and process until the obtention of the quantum metrics is achieved. This is the generalization of Corollary III.1 to the atomic space. We deduce from Eq. (69) that if K = L we have $\mathbf{U} = \mathbf{SCM}$.

Note finally that in the atomic space, occupied and virtual NTO LCAO coefficients are stored, respectively, in $\tilde{\mathbf{O}}\mathbf{O} \in \mathbb{R}^{K \times N}$ and $\tilde{\mathbf{V}}\mathbf{V} \in \mathbb{R}^{K \times (L-N)}$, where $\mathbf{O}$ and $\mathbf{V}$ are the left and right matrices implied in the SVD of $\mathbf{T}$ (see Eq. (24)).

## Author details

Thibaud Etienne

Address all correspondence to: thibaud.etienne@umontpellier.fr

Institut Charles Gerhardt, CNRS and Université de Montpellier, Montpellier, France

## References

[1] Plasser F, Lischka H. Analysis of excitonic and charge transfer interactions from quantum chemical calculations. Journal of Chemical Theory and Computation. Aug. 2012;**8**:2777-2789

[2] Guido CA, Cortona P, Adamo C. Effective electron displacements: A tool for time-dependent density functional theory computational spectroscopy. The Journal of Chemical Physics. Mar. 2014;**140**:104101

[3] Etienne T, Assfeld X, Monari A. Toward a quantitative assessment of electronic transitions charge-transfer character. Journal of Chemical Theory and Computation. Sept. 2014;**10**:3896-3905

[4] Etienne T, Assfeld X, Monari A. New insight into the topology of excited states through detachment/attachment density matrices-based centroids of charge. Journal of Chemical Theory and Computation. Sept. 2014;**10**:3906-3914

[5] Etienne T. Probing the locality of excited states with linear algebra. Journal of Chemical Theory and Computation. Feb 2015;**11**:1692-1699

[6] Guido CA, Cortona P, Mennucci B, Adamo C. On the Metric of Charge Transfer Molecular Excitations: A Simple Chemical Descriptor. Journal of Chemical Theory and Computation. July 2013;**9**:3118-3126

[7] Ciofini I, Le Bahers T, Adamo C, Odobel F, Jacquemin D. Through-Space Charge Transfer in Rod-Like Molecules: Lessons from Theory. The Journal of Physical Chemistry C. June 2012;**116**:11946-11955

[8]  Ciofini I, Le Bahers T, Adamo C, Odobel F, Jacquemin D. Correction to "through-space charge transfer in rod-like molecules: lessons from theory". The Journal of Physical Chemistry C. July 2012;**116**:14736-14736

[9]  Le Bahers T, Adamo C, Ciofini I. A Qualitative Index of Spatial Extent in Charge-Transfer Excitations. Journal of Chemical Theory and Computation. Aug. 2011;**7**:2498-2506

[10]  García G, Adamo C, Ciofini I. Evaluating push–pull dye efficiency using TD-DFT and charge transfer indices. Physical Chemistry Chemical Physics. Oct. 2013;**15**:20210-20219

[11]  Jacquemin D, Bahers TL, Adamo C, Ciofini I. What is the "best" atomic charge model to describe through-space charge-transfer excitations? Physical Chemistry Chemical Physics. Mar. 2012;**14**:5383-5388

[12]  Cèron-Carrasco JP, Siard A, Jacquemin D. Spectral signatures of thieno [3, 4-b] pyrazines: Theoretical interpretations and design of improved structures. Dyes and Pigments. Dec. 2013;**99**:972-978

[13]  Bäppler SA, Plasser F, Wormit M, Dreuw A. Exciton analysis of many-body wave functions: Bridging the gap between the quasiparticle and molecular orbital pictures. Physical Review A. Nov. 2014;**90**:052521

[14]  Plasser F, Thomitzni B, Bäppler SA, Wenzel J, Rehn DR, Wormit M, Dreuw A, Statistical analysis of electronic excitation processes: Spatial location, compactness, charge transfer, and electron-hole correlation. Journal of Computational Chemistry. 2015;**36**(21):1620-1620

[15]  Mewes SA, Plasser F, Dreuw A. Communication: Exciton analysis in time-dependent density functional theory: How functionals shape excited-state characters. The Journal of Chemical Physics. 2015;**143**(17):171101

[16]  Etienne T, Assfeld X, Monari A. QM/MM modeling of Harmane cation fluorescence spectrum in water solution and interacting with DNA. Computational and Theoretical Chemistry. Feb. 2014;**1040-1041**:367-372

[17]  Etienne T, Gattuso H, Michaux C, Monari A, Assfeld X, Perpète EA. Fluorene-imidazole dyes excited states from first-principles calculations–Topological insights. Theoretical Chemistry Accounts. 2016;**135**(4):1-11

[18]  Etienne T, Chbibi L, Michaux C, Perpète EA, Assfeld X, Monari A. All-organic chromophores for dye-sensitized solar cells: A theoretical study on aggregation. Dyes and Pigments. Feb. 2014;**101**:203-211

[19]  Duchanois T, Etienne T, Beley M, Assfeld X, Perpète EA, Monari A, Gros PC. Heteroleptic Pyridyl-Carbene Iron Complexes with Tuneable Electronic Properties. European Journal of Inorganic Chemistry. Aug. 2014;**2014**:3747-3753

[20]  Dedeoglu B, Monari A, Etienne T, Aviyente V, Ozen AS. Detection of nitroaromatic explosives based on fluorescence quenching of silafluorene-and silole-containing polymers: A time-dependent density functional theory study. The Journal of Physical Chemistry C. Sept. 2014;**118**:23946-23953

[21] Ho EK-L, Etienne T, Lasorne B. Vibronic properties of para-polyphenylene ethynylenes: TD-DFT insights. The Journal of Chemical Physics. 2017;**146**(16):164303

[22] Ehrenreich H, Cohen MH. Self-consistent field approach to the many-electron problem. Physics Reviews. Aug 1959;**115**:786-790

[23] Hirata S, Head-Gordon M, Bartlett RJ. Configuration interaction singles, time-dependent Hartree–Fock, and time-dependent density functional theory for the electronic excited states of extended systems. The Journal of Chemical Physics. 1999;**111**(24):10774-10786

[24] Hirata S, Head-Gordon M. Time-dependent density functional theory within the Tamm–Dancoff approximation. Chemical Physics Letters. Dec. 1999;**314**:291-299

[25] Dreuw A, Head-Gordon M. Single-reference ab initio methods for the calculation of excited states of large molecules. Chemical Reviews. Nov. 2005;**105**:4009-4037

[26] Casida ME. In: Seminario JM, editor. Time-dependent density functional response theory of molecular systems: theory, computational methods, and functionals. Theoretical and Computational Chemistry, Vol. 4 of Recent Developments and Applications of Modern Density Functional Theory. Amsterdam, Elsevier; 1996. p. 391-439

[27] Casida ME. Time-dependent density-functional theory for molecules and molecular solids. Journal of Molecular Structure: THEOCHEM. Nov. 2009;**914**:3-18

[28] Head-Gordon M, Grana AM, Maurice D, White CA. Analysis of electronic transitions as the difference of electron attachment and detachment densities. The Journal of Physical Chemistry. 1995;**99**:14261

[29] Plasser F, Wormit M, Dreuw A. New tools for the systematic analysis and visualization of electronic excitations. I. Formalism. The Journal of Chemical Physics. July 2014;**141**:024106

[30] Plasser F, Bäppler SA, Wormit M, Dreuw A. New tools for the systematic analysis and visualization of electronic excitations. II. Applications. The Journal of Chemical Physics. July 2014;**141**:024107

[31] Pastore M, Assfeld X, Mosconi E, Monari A, Etienne T. Unveiling the nature of post-linear response Z-vector method for time-dependent density functional theory. The Journal of Chemical Physics. 2017;**147**(2):024108

[32] Etienne T. Transition matrices and orbitals from reduced density matrix theory. The Journal of Chemical Physics. 2015;**142**(24):244103

[33] Martin RL. Natural Transition Orbitals. The Journal of Chemical Physics. Mar. 2003;**118**: 4775-4777

[34] Batista, Martin, John Wiley & Sons. Encyclopedia of Computational Chemistry; 2004. https://goo.gl/P4qqt7

[35] Mayer I. Using singular value decomposition for a compact presentation and improved interpretation of the CIS wave functions. Chemical Physics Letters. Apr. 2007;**437**:284-286

[36] Surján PR. Natural orbitals in CIS and singular-value decomposition. Chemical Physics Letters. May 2007;**439**:393-394

[37] Furche F. On the density matrix based approach to time-dependent density functional response theory. The Journal of Chemical Physics. Apr. 2001;**114**:5982-5992

[38] Luzanov AV, Zhikol OA. Electron invariants and excited state structural analysis for electronic transitions within CIS, RPA, and TDDFT models. International Journal of Quantum Chemistry. 2010;**110**(4):902-924

[39] Luzanov AV, Sukhorukov AA, Umanskii V. Application of transition density matrix for analysis of excited states. Theoretical and Experimental Chemistry. July 1976;**10**:354-361

[40] Li Y, Ullrich CA. Time-dependent transition density matrix. Chemical Physics. 2001;**391**(1):157-163

[41] Tretiak S, Mukamel S. Density matrix analysis and simulation of electronic excitations in conjugated and aggregated molecules. Chemical Reviews. Sept. 2002;**102**:3171-3212

[42] Wu C, Malinin SV, Tretiak S, Chernyak VY. Exciton scattering approach for branched conjugated molecules and complexes. III. Applications. The Journal of Chemical Physics. 2008;**129**(17):174113

[43] Tretiak S, Igumenshchev K, Chernyak V. Exciton sizes of conducting polymers predicted by time-dependent density functional theory. Physical Review B. 2005;**71**(3):033201

[44] Amos AT, Hall GG. Single determinant wave functions. Proceedings of the Royal Society of London. Series A. Mathematical and Physical Sciences. Oct. 1961;**263**:483-493

[45] Frisch M et al. Gaussian 09 Revision B.01. Wallingford, CT: Gaussian Inc.; 2009

[46] Adamo C, Barone V. Toward reliable density functional methods without adjustable parameters: The PBE0 model. The Journal of Chemical Physics. Apr. 1999;**110**:6158-6170

[47] Adamo C, Scuseria GE, Barone V. Accurate excitation energies from time-dependent density functional theory: Assessing the PBE0 model. The Journal of Chemical Physics. Aug. 1999;**111**:2889-2899

[48] Frisch MJ, Pople JA, Binkley JS. Self-consistent molecular orbital methods 25. Supplementary functions for Gaussian basis sets. The Journal of Chemical Physics. Apr. 1984;**80**:3265-3269

[49] Assfeld X, Monari A, Very T, Etienne T. Nancy-Ex 2.0 software suite. http://nancyex.sourceforge.net

[50] Etienne T. TAELES software suite. http://taeles.wordpress.com